\documentclass[aps, prl, twocolumn,showkeys,showpacs,amsmath,amssymb,superscriptaddress,notitlepage,floatfix]{revtex4-1}

\usepackage{graphicx,graphics}
\usepackage{dcolumn}
\usepackage{amsmath,amssymb,amsfonts}
\usepackage{latexsym,verbatim}
\usepackage{bm,bbold}
\usepackage{color}
\usepackage{ulem}
\usepackage[percent]{overpic}
\usepackage[breaklinks=true,colorlinks,citecolor=blue,linkcolor=blue,urlcolor=blue]{hyperref}
\usepackage[font=small,labelfont=bf]{caption}
\usepackage{subcaption}

\newcommand*\diff{\mathop{}\!\mathrm{d}}

\begin{document}
\title{Prospects for the detection of electronic pre-turbulence in graphene}
\author{A. Gabbana}
\affiliation{Universit\`a di Ferrara and INFN-Ferrara, I-44122 Ferrara,~Italy}
\affiliation{Bergische Universit\"at Wuppertal, D-42119 Wuppertal,~Germany}
\author{M. Polini}
\affiliation{Istituto Italiano di Tecnologia, Graphene Labs, Via Morego 30, I-16163 Genova,~Italy}
\author{S. Succi}
\affiliation{Center for Life Nano Science @ Sapienza, Italian Institute of Technology, 
Viale Regina Elena 295, I-00161 Roma,~Italy}
\affiliation{Istituto Applicazioni del Calcolo, National Research Council of Italy, Via dei Taurini 19, I-00185 Roma,~Italy}
\author{R. Tripiccione}
\affiliation{Universit\`a di Ferrara and INFN-Ferrara, I-44122 Ferrara,~Italy}
\author{F.M.D. Pellegrino}
\affiliation{Dipartimento di Fisica e Astronomia, Universit\`a di Catania, Via S. Sofia 64, I-95123 Catania,~Italy}
\affiliation{INFN, Sez.~Catania, I-95123 Catania,~Italy}
\begin{abstract}
Based on extensive numerical simulations, accounting for electrostatic interactions and dissipative
electron-phonon scattering, we propose experimentally realizable geometries capable of sustaining
electronic pre-turbulence in graphene samples. 
In particular, pre-turbulence is predicted to occur at experimentally attainable values of the Reynolds number 
between $10$ and $50$, over a broad spectrum of frequencies between $10$ and $100~{\rm GHz}$.
\end{abstract}
\maketitle

%
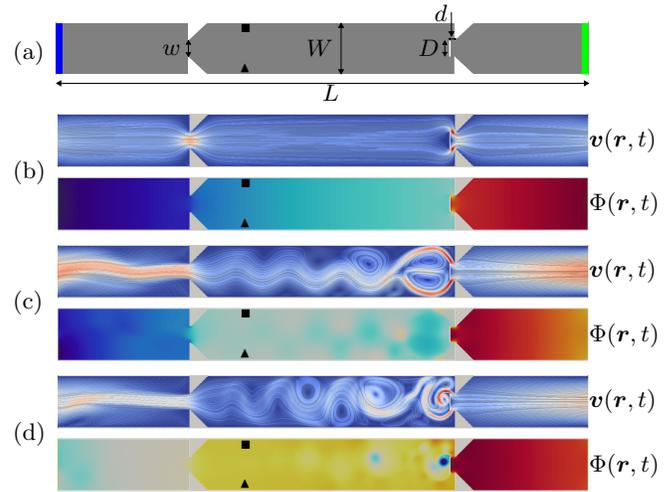
\begin{figure}[t]
  \begin{overpic}[width=.82\columnwidth]{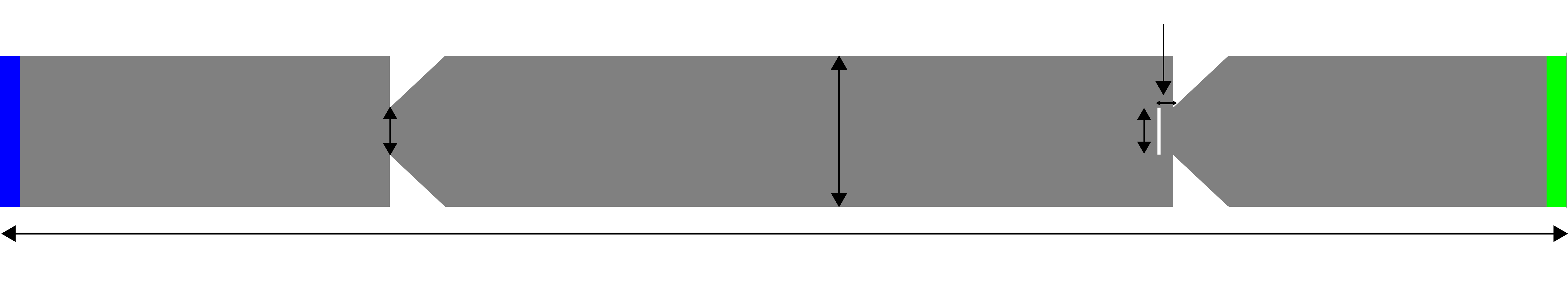}
    \put(-8, 9){(a)}
    \put(50, 1){$L$}
    \put(47, 9.5){$W$}
    \put(71,16){$d$}
    \put(68, 9.5){$D$}
    \put(20,9.7){$w$}
    \put(34.5 ,  6.4){\scalebox{.6}{$\blacktriangle$}}
    \put(34.75, 13.9){\scalebox{.5}{$\blacksquare$}}
  \end{overpic}
  \begin{overpic}[width=.82\columnwidth]{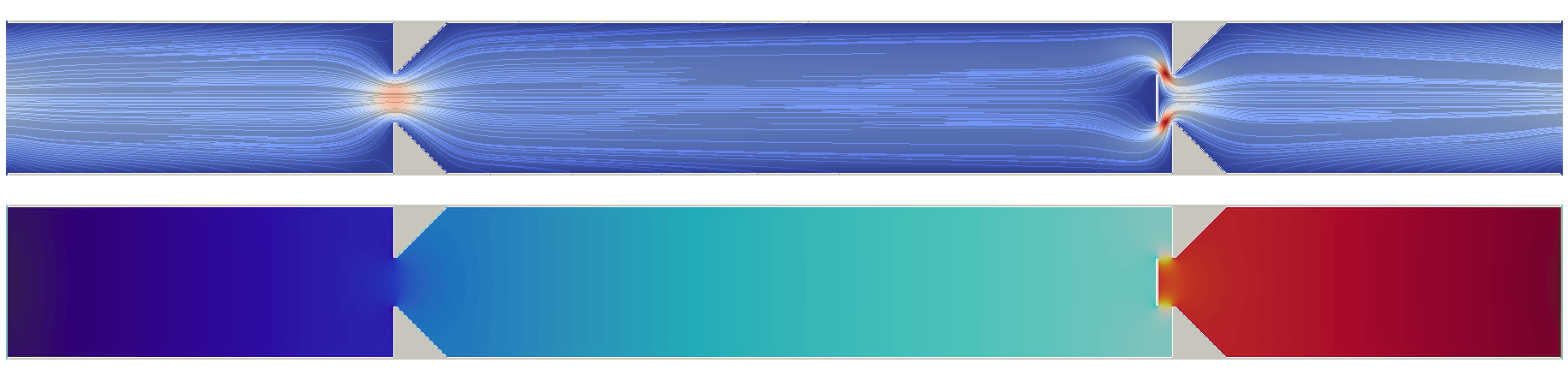}
    \put(-8,11){(b)}
    \put(100, 17){${\bm v}({\bm r}, t)$}
    \put(100, 5){$\Phi({\bm r},t)$}
    \put(34.5 ,  1.9){\scalebox{.6}{$\blacktriangle$}}
    \put(34.75,  9.3){\scalebox{.5}{$\blacksquare$}}  
  \end{overpic}
  \begin{overpic}[width=.82\columnwidth]{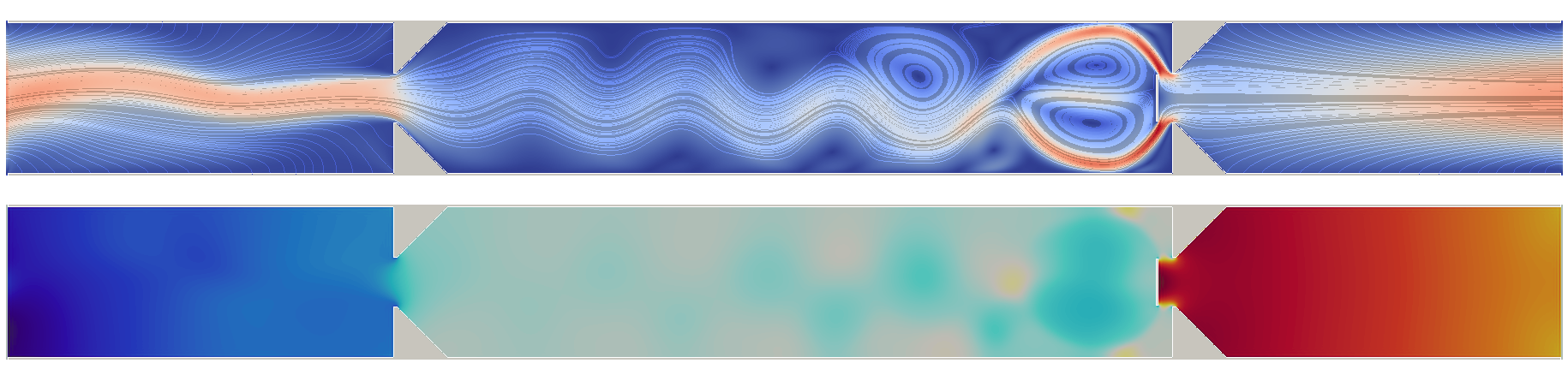}
    \put(-8,11){(c)}
    \put(100, 17){${\bm v}({\bm r}, t)$}
    \put(100, 5){$\Phi({\bm r},t)$}    
    \put(34.5 ,  1.9){\scalebox{.6}{$\blacktriangle$}}
    \put(34.75,  9.3){\scalebox{.5}{$\blacksquare$}}  
  \end{overpic}  
  \begin{overpic}[width=.82\columnwidth]{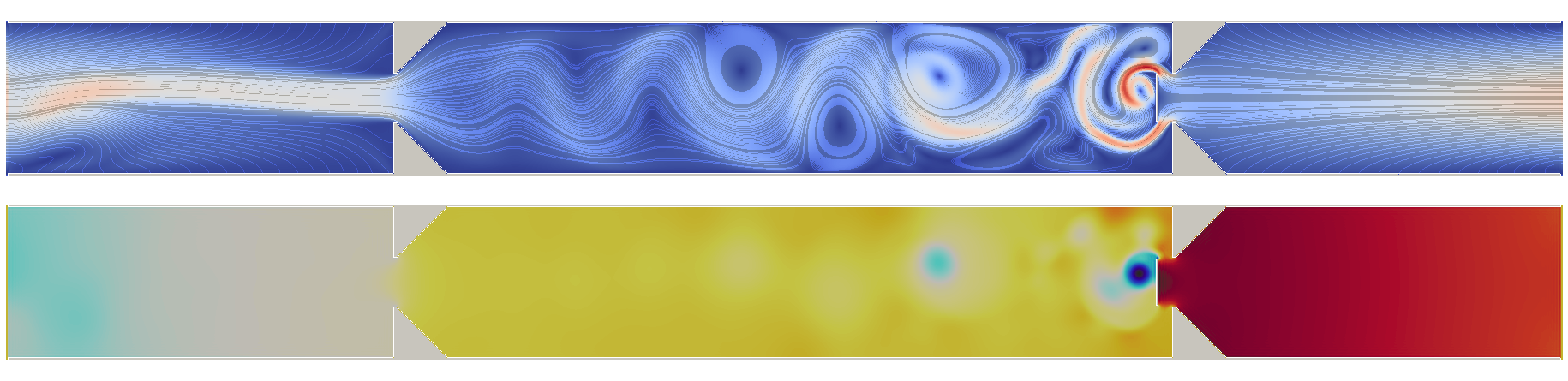}
    \put(-8,11){(d)}
    \put(100, 17){${\bm v}({\bm r}, t)$}
    \put(100, 5){$\Phi({\bm r},t)$}    
    \put(34.5 ,  1.9){\scalebox{.6}{$\blacktriangle$}}
    \put(34.75,  9.3){\scalebox{.5}{$\blacksquare$}}    
  \end{overpic}  
\caption{(Color online) Pre-turbulence in high-quality graphene. Panel a) Geometrical details of the setup analyzed in this work. 
          Two graphene leads of width $W=1~{\rm \mu m}$ are attached via ``funnels'' to a central area. 
          Current is injected through an orifice of width $w = 0.32~{\rm \mu m}$ with an obstacle of length 
          $D=0.3~{\rm \mu m}$ placed at a lateral distance $d = 0.1~{\rm \mu m}$ from the orifice. 
          Panels b)-d) Snapshots of simulations for several values of the injected current. 
          Panel b) Velocity field ${\bm v}({\bm r}, t)$ (top) and electrochemical potential 
          $\Phi({\bm r}, t)$ (bottom) for an injected current $I = 10^{-6}~{\rm A}$. 
          Panel c) Same as in panel b) but for an injected current 
          $I = 5\cdot 10^{-4}~{\rm  A}$.
          Panel d) Same as in panels b) and c) but for $I = 10^{-3}~{\rm A}$. 
          Data in panels b)-d) have been obtained by setting  
          $\nu = 4\times10^{-4}~{\rm m}^2/{\rm s}$, $\tau_{\rm D} = 50~{\rm ps}$, and 
          $C_{\rm g}/e^2 = 1.52 \cdot 10^{35}~{\rm J}^{-1} {\rm m}^{-2}$
          (See the text for definitions of all quantities).
        }\label{fig:snapshot}
\end{figure}
%
{\it Introduction.---}Hydrodynamic theory~\cite{landaufluidmechanics,falkovich_book} has proven very successful 
in describing a large variety of physical systems, across a broad range of scales, temperature and density
regimes. The ultimate reason of this success is ``universality'', namely the insensitivity of the hydrodynamic description to 
the details of the underlying microscopic physics, as long as such details do not spoil the basic mass, momentum, and energy
conservation laws, which underpin the emergence of hydrodynamic behaviour.
 
Under such conditions, at ``sufficiently large'' scales  (``large'' meaning much larger than the typical microscopic interaction length), the 
specific details of the interactions among the constituent particles do not affect the structure of the hydrodynamic 
equations, but only the actual values of the transport coefficients controlling dissipative effects, such as the shear and bulk viscosity, as well 
as the thermal conductivity. 

Even if electrons roaming in a crystal can loose energy and momentum towards impurities and the lattice, transport in systems where the mean free path for electron-electron collisions is the shortest length scale of the problem, can also be described by hydrodynamic theory and the Navier-Stokes equations~\cite{gurzhi_spu_1968,dyakonov_prl_1993,dyakonov_prb_1995,dyakonov_ieee_1996,conti_prb_1999,govorov_prl_2004,muller_prb_2008,fritz_prb_2008,muller_prl_2009,bistritzer_prb_2009,mendoza_prl_2011,svintsov_jap_2012,mendoza_scirep_2013,tomadin_prb_2013,narozhny_prb_2015,briskot_prb_2015,torre_prb_2015,levitov_naturephys_2016,pellegrino_prb_2016,principi_prb_2016,lucas_prb_2016,alekseev_prl_2016,falkovich_prl_2017,guo_pnas_2017,pellegrino_prb_2017,levchenko_prb_2017,scaffidi_prl_2017,delacretaz_prl_2017,petrov_arxiv_2018,ho_prb_2018,lucas_prb_2018_I,lucas_prb_2018_II}. Interestingly, also phonon transport is expected to display hydrodynamic features~\cite{fugallo_nanolett_2014, capellotti_natcomm_2015}.

Recent experiments carried out in high-quality encapsulated graphene sheets~\cite{bandurin_science_2016,kumar_naturephys_2017,bandurin_arxiv_2018,berdyugin_arxiv_2018} 
and GaAs quantum wells~\cite{braem_arxiv_2018} have demonstrated unique qualitative features of hydrodynamic electron transport, namely a negative quasi-local resistance~\cite{bandurin_science_2016,bandurin_arxiv_2018,berdyugin_arxiv_2018,braem_arxiv_2018} and super-ballistic electron flow~\cite{kumar_naturephys_2017}, providing, for the first time, the ability to directly measure the dissipative shear viscosity $\eta$ of a two-dimensional (2D) electron system. A different experiment~\cite{crossno_science_2016} has shown that, near charge neutrality, electron-electron interactions in graphene are strong enough to yield substantial violations of the Wiedemann-Franz law. Evidence of hydrodynamic transport has also been reported in quasi-2D channels of palladium cobaltate~\cite{moll_science_2016}. 
For a recent review, see Ref.~\onlinecite{lucas_jpcm_2018}.

Given this context, it is natural to investigate conditions under which nonlinear terms of the Navier-Stokes 
equations, which have proven unnecessary so far to explain experimental results~\cite{bandurin_science_2016,kumar_naturephys_2017,bandurin_arxiv_2018,berdyugin_arxiv_2018,braem_arxiv_2018,crossno_science_2016,moll_science_2016}, may become relevant. 

In this Letter, we identify a range of geometrical and physical parameters, in which electronic pre-turbulence can be triggered 
and sustained in experimentally realizable graphene samples, provided a substantial reduction of electron-phonon scattering 
is achieved in future experiments. 
In this context, pre-turbulence refers to a regime prior to the onset of chaos, where periodic oscillations of the velocity field can be observed, without 
necessarily exhibiting chaotic behaviour~\cite{kaplan_1979}. To this purpose, we performed extensive numerical simulations {\it taking into account} electrostatic interactions and electron-phonon scattering. In particular, we propose suitable geometries for which pre-turbulence: i) occurs at experimentally achievable values of the Reynolds number and, ii) exhibits temporal fluctuations of the electrical potential over a spectrum of frequencies 
between $10$ and $100~{\rm GHz}$.
%
%
\begin{figure}[t]
\centering
\begin{overpic}[width=.9\columnwidth]{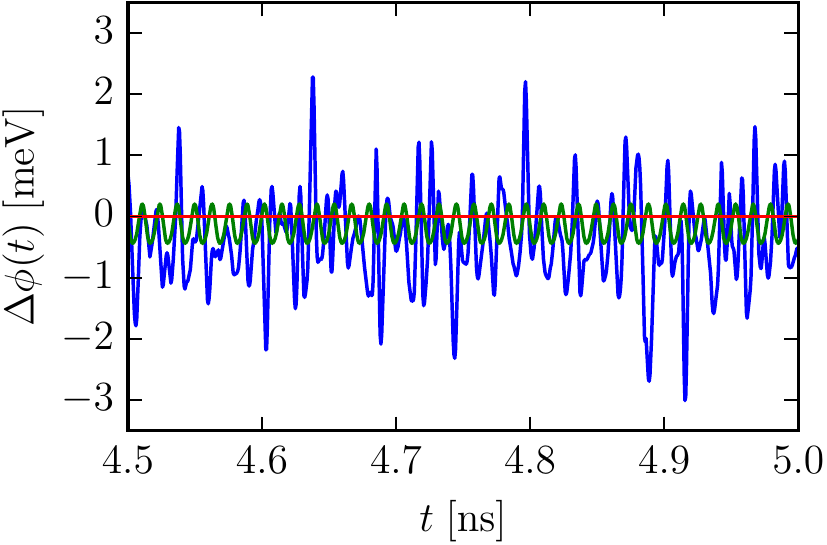}\put(0,58){(a)}
\end{overpic}
\begin{overpic}[width=.9\columnwidth]{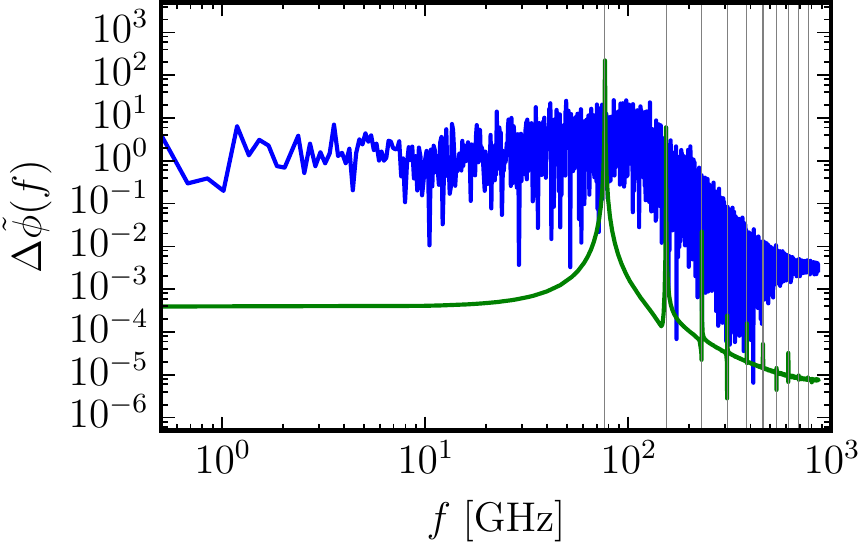} \put(0,58){(b)}
\end{overpic}
\caption{(Color online) (a) Time evolution of the electrochemical potential difference 
            $\Delta \Phi =  \Phi(\bar{\bm r}, t) - \Phi(\bar{\bm r}^{\prime}, t)$, 
            with $\bar{\bm r} = (3~\rm{\mu m}, 0.1~\rm{\mu m}) $ and 
            $\bar{\bm r}^{\prime} = (3~\rm{\mu m}, 0.9~\rm{\mu m}) $. These two points have been marked in Fig~\ref{fig:snapshot}a by a triangle ($\bar{\bm r}$) and a square ($\bar{\bm r}^{\prime}$).
            Numerical results shown is this figure have been taken from simulations using $\nu = 4\times 10^{-4}~{\rm m}^2/{\rm s}$,
            $\tau_{\rm D} = 50~{\rm ps}$, $C_{\rm g}/e^2 = 1.52 \cdot 10^{35}~{\rm J}^{-1} {\rm m}^{-2}$,
            and the following values of the injected current: $I = 10^{-6}~{\rm A}$ (red), 
            $I = 5\cdot10^{-4}~{\rm A}$ (green), and $I = 10^{-3}~{\rm A}$ (blue).
            (b) Fourier transform of the signals shown in panel (a). The gray vertical lines 
            represent the first ten harmonics of the dominant frequency of the periodic signal 
            obtained from the simulation 
            at injected current $I = 5\cdot 10^{-4}~{\rm A}$.          
          }\label{fig:fft}
\end{figure}

{\it Kinetic description and Boltzmann equation.}---The direct solution of the Navier-Stokes equations presents a numerically challenging task. 
In the last decades, it has become apparent that a broad class of complex flows can be addressed by solving suitably simplified lattice 
versions of Boltzmann's kinetic equation~\cite{succi_epl_2015} (for details see Supplementary Material).
  
For the specific 2D electron system of interest in this work, Boltzmann's kinetic equation reads as follows:
\begin{align}\label{eq:be}
  \left( \frac{\partial}{\partial t} + \frac{\bm{p}}{m} \cdot \nabla + \bm{F} \cdot \frac{\partial}{\partial \bm{p}} \right) f = \Omega
\end{align}
where $f(\bm{r},\bm{p},t)$ is the one-particle distribution function expressing the average number of particles in a small element of phase-space centered at position $\bm{r}$ with momentum $\bm{p}$ at time $t$.
In the above, $m$ is a suitable effective mass, $\bm{F}$ is the sum of all external forces acting on the system 
and $\Omega$ is the collision operator, commonly replaced by a relaxation term towards local equilibrium~\cite{BGK}.

It is well known that hydrodynamics emerges from Eq.~(\ref{eq:be}) in the limit of small Knudsen numbers~\cite{CE}, leading 
to the continuity, Navier-Stokes, and energy conservation equations. 
Microscopic details are reflected by the transport coefficients. 

The bulk viscosity $\zeta$ is negligibly small for electrons in graphene~\cite{principi_prb_2016} and while the lattice
Boltzmann equation usually features a non-zero value, it has no effect on the physics discussed here since the flow
is nearly-incompressible. 
The shear viscosity $\eta$, on the other hand, plays a crucial role~\cite{bandurin_science_2016,kumar_naturephys_2017,bandurin_arxiv_2018,berdyugin_arxiv_2018} 
and consequently it is taken in full account.
 
For the specific case of 2D electrons in doped graphene, the total force is taken in the following form:
\begin{equation}
  {\bm F} =  e\nabla \varphi({\bm r},t)  - \frac{n({\bm r}, t)v({\bm r},t)}{\tau_{\rm D}}~.
\end{equation}
The first term at the right-hand side describes electrical forces acting on a fluid element, $-e$ being the elementary charge and $\varphi({\bm r},t)$ the electrical potential in the 2D plane where electrons move.
The second term describes forces that dissipate electron momentum, i.e.~due to collisions between electrons and external agents, such as acoustic phonons in graphene. 
These are parametrized as an external friction, with a single time scale, i.e.~the Drude-like scattering time $\tau_{\rm D}$. 
This simple parametrization has proven extremely successful in describing experiments in the linear-response regime~\cite{bandurin_science_2016,kumar_naturephys_2017,bandurin_arxiv_2018,berdyugin_arxiv_2018,braem_arxiv_2018}.

Following Ref.~\onlinecite{tomadin_prb_2013}, we utilize the local capacitance approximation in which the electrical potential is approximated as
$\varphi({\bm r}, t) \approx - e \delta n({\bm r}, t)/C_{\rm g}$, where $C_{\rm g}$ is the geometrical capacitance of the graphene device of interest and $\delta n({\bm r}, t)=n({\bm r}, t)-\bar{n}$, ${\bar n}$ being the uniform value of the electron density set by a nearby metallic gate. Using a similar local approximation for the gradient of the pressure~\cite{Giuliani_and_Vignale}, i.e.~$\nabla P\approx (\partial P/\partial n)_{n({\bm r},t)\to {\bar n}}\nabla \delta n({\bm r},t)$ we can define the electrochemical potential as $\phi({\bm r},t)\equiv -e \delta n({\bm r}, t)(C^{-1}_{\rm g} + C^{-1}_{\rm Q})$, $C_{\rm Q}=2 \bar{n} e^2/E_{\rm F}$ being the so-called quantum capacitance~\cite{Giuliani_and_Vignale} and $E_{\rm F}=\hbar v_{\rm F}\sqrt{\pi {\bar n}}$ the Fermi energy in single-layer graphene (SLG). Finally, $v_{\rm F} \simeq 10^{6}~{\rm m}/{\rm s}$ is the Fermi velocity of massless Dirac fermions in SLG. With reference to Eq.~(\ref{eq:be}), we use the usual effective mass $m=E_{\rm F}/v^2_{\rm F}$ for SLG.

Our numerical results are based on extensive numerical simulations 
of the geometry shown in Fig.~\ref{fig:snapshot}a, which can be easily realized experimentally with current technology, and 
for a large set of values of the relevant physical parameters (see Tab.~\ref{tab:parameters}). 
All cases considered in this work fall in a regime of very small Mach number ${\rm Ma}$, in which compressibility effects can safely be neglected.

The Mach number is defined as the ratio between the plasma-wave velocity $v_{\rm PW}$ and the fluid velocity of the electron fluid, with 
$v_{\rm PW}=\sqrt{e^2 \bar{n} v_{\rm F}^2/(C E_{\rm F})}$, where $C^{-1}=C_{\rm g}^{-1}+C_{\rm Q}^{-1}$. For the device geometry shown in Fig.~\ref{fig:snapshot}a and the parameters used in all our simulations, ${\rm Ma}\ll 1$. (This has been explicitly verified a posteriori for all cases. For example, for the simulations corresponding to 
Figs.~\ref{fig:snapshot}(b-d), we have ${\rm Ma} \approx 0.0015$, $0.08$, and, $0.12$, respectively.) A small value of ${\rm Ma}$ in turn implies the quasi-incompressibility of the electron fluid. 
As mentioned earlier on, in this regime we have resorted to a Lattice Boltzmann (LB) approach~\cite{Succi_book}, which, among others, offers the 
 advantage of a comparatively simple handling of non-idealized geometrical boundary conditions. In this work, we use a non-relativistic LB scheme, since relativistic approaches~\cite{rlbm1,rlbm2,rlbm3} are appropriate only very close to the charge neutrality point, where charge and energy flows are coupled~\cite{lucas_jpcm_2018}. Technical details on this numerical approach are reported in Ref.~\onlinecite{SOM}. 
%
\begin{table}[t]
\caption{ Typical values of physical parameters of state of the art experiments compared with those used in our simulations. 
Refer to Fig.~\ref{fig:snapshot}a for the definition of $L$ and $W$. All other parameters are defined in the main text. \label{tab:parameters}}
\centering
\begin{tabular}{l|c|c}
%
      & Typical experiments & This work \\
\hline
$L               $ & $             5\sim30~\rm{[\mu m]}        $ & $               10~\rm{[\mu m]}         $ \\
$W               $ & $           1\sim5   ~\rm{[\mu m]}        $ & $                1~\rm{[\mu m]}         $ \\
$\bar{n}         $ & $  0.5\sim 4 \cdot 10^{12}~\rm{[cm^{-2}]} $ & $   2\cdot10^{12}~\rm{[cm^{-2}]}        $ \\
$I               $ & $      10^{-3}\sim1 ~\rm{[mA]}            $ & $       10^{-3}\sim1 ~\rm{[mA]}         $ \\
$\nu             $ & $      0.01\sim0.1~\rm{[m^2 / s]}         $ & $  10^{-4}\sim10^{-3}~\rm{[m^2 / s]}    $ \\
$\tau_{\rm D}    $ & $              1\sim5~\rm{[ps]}           $ & $            1\sim400~\rm{[ps]}         $ \\
$C_{\rm g} / e^2 $ & $3.03\cdot 10^{34}~\rm{[J^{-1} m^{-2}]}   $ & $3.03\cdot 10^{35}~\rm{[J^{-1} m^{-2}]} $ \\
\hline
\end{tabular}
\end{table}
%

{\it Numerical results.---}We consider a geometry close to the one used in recent experimental work~\cite{kumar_naturephys_2017}, which made use of a constriction to emphasize a clear crossover from the ballistic Sharvin regime to the hydrodynamic regime as a function of temperature. 
Such geometry is sketched in Fig.~\ref{fig:snapshot}a, with the addition of a thin linear obstacle, placed in front 
of the constriction, with the intent of triggering pre-turbulent regimes at low Reynolds numbers.

Fig.~\ref{fig:snapshot} qualitatively summarizes our finding. For appropriate values of the transport parameters (low enough kinematic viscosity $\nu = \eta/(n m)$ and large enough $\tau_{\rm D}$) a laminar behaviour is found for low values of the current ($10^{-3}~\rm{mA}$, Fig.~\ref{fig:snapshot}b) injected in the sample. As the value of the injected current is increased ($0.5 - 1.0~\rm{mA}$, Fig.~\ref{fig:snapshot}c/d, and, correspondingly, the typical fluid element velocity increases), a transition to a pre-turbulent behaviour takes place 
(identified with a procedure described later in the text).

Present-day experiments cannot map the fluid velocity everywhere in the sample, but typically can only measure the 
electrochemical potential (also mapped in Fig.~\ref{fig:snapshot}) at selected sites on the boundaries. 

The expected result of such measurements is shown in Fig.~\ref{fig:fft}a, displaying the electrochemical potential difference between 
locations corresponding to the black square and triangle in Fig.~\ref{fig:snapshot}a; here again, we appreciate 
a clear change from a constant to a periodic, to a more irregular trend, which is best analyzed in the frequency domain, see Fig.~\ref{fig:fft}b.

\begin{figure}[t]
\centering
\includegraphics{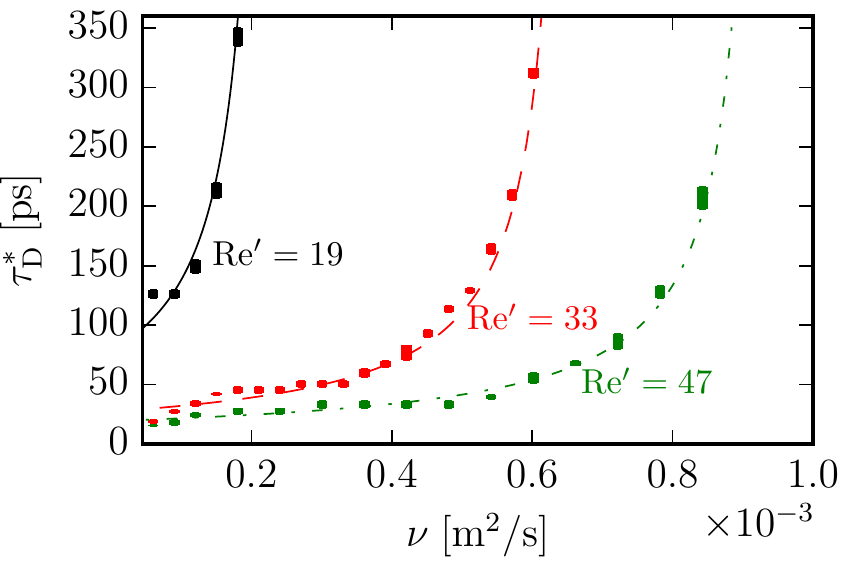}
\caption{(Color online) 
         Critical value $\tau^*_{\rm D}$ of the Drude-like scattering time as a function of the kinematic 
         viscosity $\nu$, for which a transition from a laminar to a pre-turbulent regime is observed. 
         Thick vertical bars represent results of numerical simulations (refer to Ref.~\onlinecite{SOM} 
         for details on how these intervals are established), with the following values of the injected 
         current: $I = 10^{-4}~\rm{A}$ (black), $I = 5\cdot10^{-4}~\rm{A}$ (red), $I = 10^{-3}~\rm{A}$ (green). 
         Lines represent iso-Reynolds curves, where ${\rm Re}^{\prime}$ as in Eq.~(\ref{eq:reyn}) is used in 
         the definition of a Reynolds number that includes extrinsic dissipation due to $\tau_{\rm D}$ and 
         $\ell$ is a fitting parameter. Lines represents fits to the numerical data: $I = 10^{-4}~\rm{A}$ 
         (solid black line), $I = 5\cdot10^{-4}~\rm{A}$ (red dashed line), and $I = 10^{-3}~\rm{A}$ 
         (green dash-dotted line). Refer to Fig~\ref{fig:snapshot}a for details on the geometry used in the simulations. 
         Numerical data in this figure have been obtained by setting 
         $C_{\rm g}/e^2 = 3.03 \cdot 10^{35}~ {\rm J}^{-1} {\rm m}^{-2}$.
         }\label{fig:nu-vs-tauD}
\end{figure}

The present simulations cover a wide region in the $\nu$-$\tau_{\rm D}$ plane. 
Results are collected in Fig.~\ref{fig:nu-vs-tauD}, showing the smallest value of $\tau_{\rm D}$ as a function of $\nu$, for which a transition to an observable pre-turbulent regime occurs, denoted by the symbol $\tau^*_{\rm D}$.

Points in Fig.~\ref{fig:nu-vs-tauD} refer to experimentally achievable values of the injected current of the order of $\approx 1~{\rm mA}$. 
They have been determined using the onset of a transverse current along the middle section of the device as a discriminating factor;
the upper end of these points are simulations for which the root mean square of the transverse current exceeds $1 \%$ of the 
magnitude of the injected current (more details in the Supplementary Material).

Recent works~\cite{bandurin_science_2016,kumar_naturephys_2017} have reported direct experimental measurements of the kinematic viscosity $\nu$ of the 2D electron system in graphene, which are on the order of $\nu \lesssim~0.1~{\rm m^2/s}$. As far as electron-phonon interactions are concerned, state-of-the-art experiments in graphene encapsulated between hexagonal Boron Nitride (hBN) crystals display $\tau_{\rm D}$ ranging between $1$ and $2~\rm{ps}$ in the temperature range $70$-$300~{\rm K}$, where hydrodynamic behaviour is strongest.
Inspection of Fig.~\ref{fig:nu-vs-tauD} may therefore convey disappointing news: for values of the parameters currently 
achieved in experiments, no pre-turbulent behaviour can be detected.
The mitigating observation is that substantial, but not unconceivable, improvements of the transport 
parameters may eventually turn the picture for good. For example, the viscosity of the electron liquid at elevated injection currents, as those needed to achieve the pre-turbulent regime, is expected to be much smaller than that in the linear-response regime, due to Joule heating~\cite{deJong_prb_1995}, which notably increases the electron temperature above the lattice temperature. 
Moreover, recent material science advances~\cite{stampfer}, have enabled much larger values of $\tau_{\rm D}$ than  those measured in hBN-encapsulated graphene. Such large values of $\tau_{\rm D}$ can be obtained by using different encapsulating materials, such as ${\rm WSe}_{2}$, which are currently believed to quench scattering of electrons against acoustic phonons in graphene~\cite{stampfer}.

A further encouraging result is that the frequency distribution of the electro-chemical potential falls within a measurable regime, 
if only with suitably designed experiments.

From a purely fluid-dynamics point of view, it may be interesting to characterize the crossover line clearly shown in Fig.~\ref{fig:nu-vs-tauD} in terms of an appropriate figure of merit. To this purpose, we develop a simplified model, whose starting point is the role played by the Reynolds number as an indicator of turbulence. In the present case, the turbulence-suppressing effect of the dissipative term in the Navier-Stokes equation is augmented by electron-phonon scattering. On purely dimensional grounds, it proves expedient to introduce a modified Reynolds number ${\rm Re^{\prime}}$, incorporating
 the effect of electron-phonon dissipation, namely: 
\begin{equation}\label{eq:reyn} 
  {\rm Re^{\prime}} = \frac{|{\bm v}|~\ell}{\displaystyle \nu + \frac{\ell^2}{\tau_{\rm D}}} \quad ,
\end{equation}
with $|{\bm v}|$ a typical fluid-element velocity and $\ell$ a typical length scale for the system at hand.

This very simple model proves adequate to characterize the actual behaviour of the system. 
Lines in Fig.~\ref{fig:nu-vs-tauD} are level lines for ${\rm Re}'$, which capture the trend of the different datasets. 
In Eq.~(\ref{eq:reyn}), we use the inlet velocity and obtain $\ell = 0.135~\rm{\mu m}$
through a linear fit. Such value turns out to be pretty close to the typical geometrical features of the simulated layout.

We obtain the following estimates for the critical modified Reynolds numbers: ${\rm Re^{\prime}} \sim 19$ for $I = 10^{-4}~\rm{A}$,
${\rm Re^{\prime}} \sim 33$ for $I = 5\cdot10^{-4}~\rm{A}$ and ${\rm Re^{\prime}} \sim 47$ for $I = 10^{-3}~\rm{A}$. 

We do not wish to attach any deep meaning to this parametrization, but simply note that 
it discloses a simple theoretical interpretation of the numerical results.

{\it Closing remarks.---}Summarizing, based  on extensive numerical simulations, accounting for electrostatic and dissipative effects 
due to electron-phonon scattering in experimentally realistic geometries, we have identified parameter regimes
under which electronic pre-turbulence may eventually be detected by future experiments.
To this purpose, such experiments should operate at lower levels of electron-phonon scattering (i.e.~$\tau_{\rm D} \sim 20$-$50$~\rm{ps}) than those that can be achieved in hBN-encapsulated graphene, which is possible by using different encapsulating materials~\cite{stampfer}. 
As a typical signature of electronic pre-turbulence, we predict electrical potential fluctuations in the frequency range 
between $10$ and $100~{\rm GHz}$, which should be detectable by suitably designed experiments.

We emphasize that the placement of a thin plate across the mainstream electron flow in a constricted channel proves instrumental 
in lowering the critical Reynolds number at which pre-turbulence occurs. 
Further optimization may result from a concerted effort between future numerical and experimental investigations. 

{\it Acknowledgments.---}We wish to thank Andre Geim and Iacopo Torre for useful discussions. 
A.G. has been supported by the European Union's Horizon 2020 research and innovation programme under 
the Marie Sklodowska-Curie grant agreement No.~642069. M.P. is supported by the European Union's Horizon 
2020 research and innovation programme under grant agreement No.~785219 - GrapheneCore2.
S.S. acknowledges funding from the European Research Council under the European
Union's Horizon 2020 framework Programme (No. P/2014-2020)/ERC Grant Agreement No. 739964 (COPMAT).
The numerical work has been performed on the COKA computing cluster at Universit\`a di Ferrara. 

%

\pagebreak
\widetext
\begin{center}
\textbf{\large Supplementary Information for \\``Prospects for the detection of electronic pre-turbulence in graphene''}
\end{center}
\setcounter{equation}{0}
\setcounter{figure}{0}
\setcounter{table}{0}
\setcounter{page}{1}
\makeatletter
\renewcommand{\theequation}{S\arabic{equation}}
\renewcommand{\thefigure}{S\arabic{figure}}
\renewcommand{\bibnumfmt}[1]{[S#1]}
\renewcommand{\citenumfont}[1]{S#1}

%
\section{Numerical Method}
In this section we provide a brief introduction to the Lattice Boltzmann Method (LBM), which has been 
used to carry out the numerical work presented in the main text. For a thorough introduction to LBM
the interested reader in is kindly referred to \cite{Succi_book,Krueger_book}.

Lattice Boltzmann methods are a class of numerical fluid-dynamics solvers, 
initially developed to study quasi-incompressible isothermal fluids \cite{higuera-1989,chen-1992,qian-1993}, and then 
improved  to incorporate e.g. thermo-hydrodynamical fluctuations \cite{philippi-2006,sbragaglia-2009,chikatamarla-2009}, 
or covering a wider range of fluid velocities from low-velocity to ultra-relativistic regimes \cite{mendoza-2010a,rlbm1,rlbm2}.
At variance with methods that discretize the Navies-Stokes equations, LBM stems from the mesoscopic Boltzmann equation:
\begin{align}\label{eq:be}
  \left( \frac{\partial}{\partial t} + \frac{\bm{p}}{m} \cdot \nabla + \bm{F} \cdot \frac{\partial}{\partial \bm{p}} \right) f = \Omega(f)
\end{align}
where $f(\bm{r},\bm{p},t)$ is the one-particle distribution function expressing the average number of particles in a small element of phase-space centered at position $\bm{r}$ with momentum $\bm{p}$ at time $t$.
In the above, $m$ is a suitable effective mass, $\bm{F}$ is the sum of all external forces acting on the system.
The collisional operator $\Omega(f)$, describing the changes in $f$ due to particle collisions, is 
commonly replaced by the single-time relaxation BGK model \cite{bhatnagar-1954}:
\begin{equation}\label{eq:bgk}
  \Omega(f) = \frac{1}{\tau} \left( f^{eq} - f \right) \quad .
\end{equation}
Using this model the evolution of the system is described by a relaxation process, with 
relaxation time $\tau$, towards a local equilibrium $f^{eq}$ given by the Maxwell-Boltzmann distribution:
\begin{equation}\label{eq:mb-distribution}
  f^{eq} =   n \left( \frac{1}{2 \pi k_B}  \frac{ m }{ T } \right) 
                       \exp{ \left( - \frac{1}{2 k_B } \frac{ m }{ T } (\bm{\xi} - \bm{u})^2  
                \right) } \quad .
\end{equation}
Macroscopic quantities like the particle number density $n(\bm{r}, t)$, velocity $\bm{u}(\bm{r}, t)$ and temperature $T(\bm{r}, t)$ are linked to the microscopic 
velocity ($\bm{\xi}$) moments of $f$:
\begin{align}
                    n (\bm{r}, t) &=               \int f(\bm{r}, \bm{\xi}, t)                                    \diff \bm{\xi} \label{eq:mb-first-moment}   \\
  n (\bm{r}, t) \bm{u}(\bm{r}, t) &=               \int f(\bm{r}, \bm{\xi}, t)   \bm{\xi}                         \diff \bm{\xi} \label{eq:mb-second-moment}  \\
  n (\bm{r}, t) T(\bm{r}, t)      &= \frac{1}{2} m \int f(\bm{r}, \bm{\xi}, t) | \bm{\xi} - \bm{u}(\bm{r}, t) |^2 \diff \bm{\xi} \label{eq:mb-third-moment} \quad . 
\end{align}

In the derivation of his 13-moments method, Grad \cite{grad-1949,grad-1949b} made an important observation on the link between the Maxwell-Boltzmann distribution and the Hermite polynomials. In fact, by expanding the equilibrium distribution
\begin{equation}\label{eq:feq-expansion}
  f^{eq}(\bm{r}, \bm{\xi}, t ) = \omega( \bm{\xi}) \sum_{k = 0}^{\infty} \frac{1}{k!} a^{(k)}(\bm{r}, t) H^{(k)} (\bm{\xi}) \quad ,
\end{equation}
with $a^{(k)}$ the projection coefficients
\begin{equation}\label{eq:projection-coefficients}
  a^{(k)}(\bm{r}, t) = \int f^{eq}(\bm{r}, \bm{\xi}, t )  H^{(k)}(\bm{\xi}) \diff \bm{\xi} \quad ,
\end{equation}
and the weighting function $\omega(\bm{\xi})$
\begin{equation}\label{eq:hermite-weight-function}
  \omega(\bm{\xi}) = \frac{1}{2 \pi} \exp{ \left( - \frac{1}{2} \bm{\xi}^2 \right) } \quad ,
\end{equation}
it is possible to show that the hydrodynamic variables can be expressed in terms of the
low-order Hermite expansion coefficients.
%
%
%
The mathematical foundation of the LBM lies on the observation that the Hermite coefficients can be calculated {\it exactly} using a Gauss-Hermite quadrature formula,
which allows to replace the (continuum) velocity space with a (small) set of discrete velocities $ \mathcal{V} = \{ \bm{e}_i \in \mathbb R^2 \}$
(refer to \cite{philippi-2006,shan-2016} for the mathematical details).

In this work we have used a iso-thermal version of the D2Q37 \cite{philippi-2006,sbragaglia-2009}, a fourth-order model, 
where the order of a model corresponds to the highest retained moment. The stencil is shown in Fig.~\ref{fig:stencil}, while
in Tab.~\ref{tab:stencil} we detail the velocity vectors and the weights of the quadrature.

\begin{minipage}{\textwidth}
  \begin{minipage}[T]{0.49\textwidth}
    \centering
    \includegraphics[width=.49\textwidth]{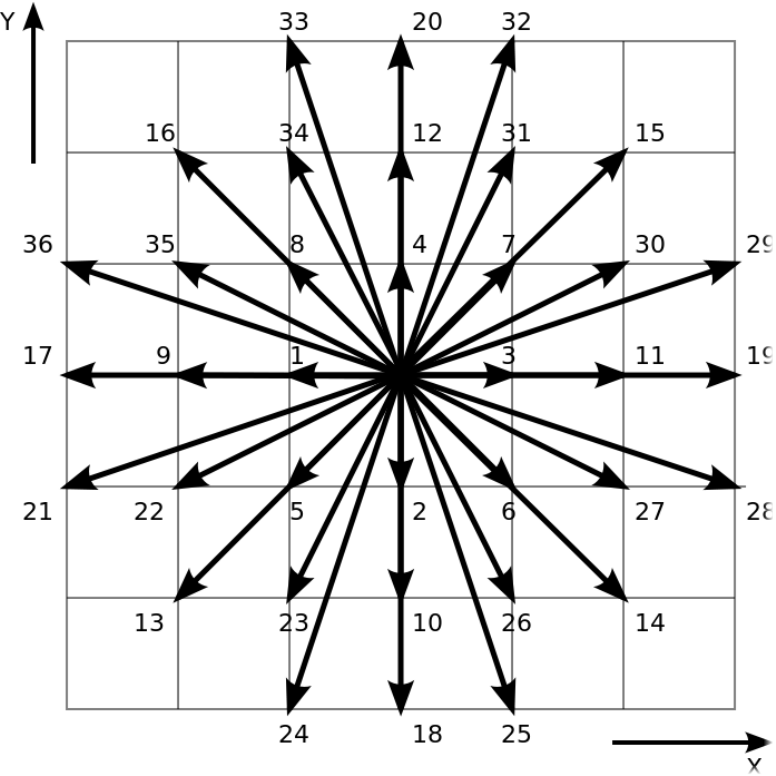}
    \captionof{figure}{Stencil for the D2Q37 model used in the simulations of
the main text. Based on the Hermite-Gauss quadrature
\cite{philippi-2006,shan-2016}, the D2Q37 can be regarded as the minimal on grid
square lattice {\it exactly} recovering the moments of the distribution  up to
the fourth order.}\label{fig:stencil}
  \end{minipage}
  \hfill
  \begin{minipage}[T]{0.49\textwidth}
    \centering
    \vspace*{1.3cm}
    \begin{tabular}{ c c }
    \hline
    $\bm{e}_i$ & $w_i$ \\
    \hline
    $\left( \phantom{\pm} 0, \phantom{\pm} 0 \right)_{\rm{\phantom{FS}}}$ & $0.2331506691323525$ \\
    $\left( \phantom{\pm} 0,          \pm  1 \right)_{\rm{FS}}          $ & $0.1073060915422190$ \\
    $\left(          \pm  1,          \pm  1 \right)_{\rm{FS}}          $ & $0.0576678598887948$ \\
    $\left( \phantom{\pm} 0,          \pm  2 \right)_{\rm{FS}}          $ & $0.0142082161584507$ \\
    $\left(          \pm  1,          \pm  2 \right)_{\rm{FS}}          $ & $0.0053530490005137$ \\
    $\left(          \pm  2,          \pm  2 \right)_{\rm{FS}}          $ & $0.0010119375926735$ \\
    $\left( \phantom{\pm} 0,          \pm  3 \right)_{\rm{FS}}          $ & $0.0002453010277577$ \\
    $\left(          \pm  1,          \pm  3 \right)_{\rm{FS}}          $ & $0.0002834142529941$ \\
    \hline 
    $ c_s$   & 0.835436007136204 \\
    \hline 
    \vspace*{-0.5cm}
    \end{tabular}
    \captionof{table}{ Quadrature weights associated to each velocity group of the D2Q37 stencil. 
                       The weights are given with 16 digits to ensure that integrals in Eq.\ref{eq:moments-quadrature} and
                       Eq.\ref{eq:discrete-feq} are  correctly computed at machine precision.
                       Here FS stands for full-symmetric meaning that, for example, $\left( 0, \pm  1 \right)_{\rm{FS}}$
                       corresponds to the velocity vectors $\{ (0, 1), (0, -1), (1, 0), (-1, 0) \}$.
                       The lattice constant $c_s$ is commonly referred to as the speed of sound in the lattice.}\label{tab:stencil}
  \end{minipage}
\end{minipage}


{\it Computational Scheme.---} For each time step and for each grid site the following 
operations are performed (see Fig.~\ref{fig:stencil} and Tab.~\ref{tab:stencil} 
for the definition of the stencil velocities $\bm{e}_i$ and the quadrature weights $w_i$, $i = 0, 1, \dots 36$ ):

\begin{enumerate}
  \item Compute the macroscopic quantities such as density and momentum:
  \begin{equation}\label{eq:moments-quadrature}
    \begin{aligned}
    n         &=  \sum_{i = 0}^{36} f_i          \\
    n \bm{u}  &=  \sum_{i = 0}^{36} f_i \bm{e}_i
    \end{aligned}
  \end{equation}
  \item Compute the equilibrium distribution:
  \begin{equation}\label{eq:discrete-feq}
    f_i^{eq} = w_i \sum_{k = 0}^{4} a^{(k)} H^{(k)} (\bm{e}_i) \quad .
  \end{equation}  

  \item Evolve the discrete Lattice Boltzmann equation:
  \begin{equation}\label{eq:bgk-lbe}
    f_i(\bm{r} + \bm{e}_i \Delta t , \bm{e}_i, t + \Delta t) - f_i(\bm{r}, \bm{e}_i, t) 
    =  
    \frac{\Delta t}{\tau} \left( f_i^{\rm{eq}}(\bm{r}, \bm{e}_i, t) - f_i(\bm{r}, \bm{e}_i, t) \right) + F^{\rm{ext}}_i \quad ,
  \end{equation}
  where $F^{\rm{ext}}_i$ is the discrete counterpart of the total external force defined in the main text.

\end{enumerate}

{\it The Chapman Enskog expansion: from from lattice Boltzmann to Navier-Stokes.---} Hydrodynamics emerges from Boltzmann's kinetic theory 
in the limit of vanishing Knudsen numbers, where the Knudsen number $\rm Kn$ is defined as the ratio between 
the molecular mean free path and the typical macroscopic length scale. It is therefore natural to think of an expansion of 
the kinetic equations in powers of a vanishingly small Knudsen number. Such asymptotic analysis can be performed using the Chapman Enskog (CE) expansion.
The CE expansion is commonly employed to show that the lattice formulation correctly recovers the Navier-Stokes equations:
%
%
\begin{align}\label{eq:ns}
  \partial_t n(\bm{r},t)  + \partial_{r_{\alpha}} \left( n(\bm{r},t) u_{\alpha}(\bm{r},t) \right) &= 0 \\
  \partial_t \left( m n(\bm{r},t) u_{\beta}(\bm{r},t) \right) + m n(\bm{r},t) u_{\beta}(\bm{r},t) \partial_{r_{\alpha}} \left( u_{\alpha}(\bm{r},t) \right) &= m n(\bm{r},t) F_{\beta}(\bm{r},t)  + \partial_{r_{\alpha}} \sigma_{\alpha \beta} \quad .
\end{align}
with the stress tensor $\sigma_{\alpha \beta}$ given by
\begin{align}\label{eq:stress-tensor}
  \sigma_{\alpha \beta} = \eta (\partial_{\alpha} u_{\beta} + \partial_\beta u_{\alpha} - \partial_{\gamma} u_{\gamma} \delta_{\alpha \beta}  ) + \zeta \partial_{\gamma} u_{\gamma} \delta_{\alpha \beta}
\end{align}
where $\eta$ and $\zeta$ are respectively the shear and the bulk viscosity.
In the above, Greek subscripts run over spatial dimensions.
The closure of the CE expansion provides the expression of the transport coefficients
connecting the microscopic and macroscopic levels. In this work we are mainly interested in the kinematic viscosity:
\begin{equation}
  \nu = \frac{\eta}{n~m} = c_s^2 \left( \tau - \frac{\Delta t}{2}\right)
\end{equation}
with $c_s$ a lattice constant (see Tab.~\ref{tab:stencil}).
Full details of the CE expansion for the D2Q37 model are reported as Appendix in \cite{scagliarini-2010}.

\section*{Parameter matching}
In an experimental perspective, we are interested in taking measurements of the electrochemical potential. Since this quantity is not a direct observable
of the lattice formulation, we need to perform a parameters matching procedure. In the main text we have defined the electrochemical potential as
\begin{equation}
 \phi= \varphi(\bm{r}) - \frac{\delta P(\bm{r})}{e \bar{n}}~,
\end{equation}
where $\delta P(\bm{r})=P(\bm{r})-\bar{P} \approx \frac{\epsilon_{\rm F}}{2} \delta n(\bm{r})$.
By employing the local capacitance approximation, $\varphi(\bm{r}) \approx - e \delta n(\bm{r})/C_{\rm g}$, 
it is simple to show that an approximation for $\phi$ is given by:
\begin{equation}\label{eq:CQ}
  \phi= -e \delta n(\bm{r}) \left( \frac{1}{C_{\rm g}} + \frac{1}{C_{\rm Q}} \right)~,
\end{equation}
where $1/C_{\rm Q}=\epsilon_{\rm F}/(2 \bar{n} e^2)$.

As described above, we use a Maxwell-Boltzmann distribution within the LBM formulation.
For this reason, it follows that the hydrostatic contribution to the electrochemical potential gives an 
{\it effective quantum capacitance} that can be written as
\begin{equation}
 \frac{1}{C_{\rm Q,MB}}=\frac{k_{\rm B} T}{\bar{n} e^2 }~.
\end{equation}
In the numerical scheme, used to describe a iso-thermal dynamic, the temperature appears only in this term.
Therefore, using the temperature as an effective parameter, we can match the correct expression for the electrochemical potential:
\begin{equation}
 k_{\rm B} T= \frac{\epsilon_{\rm F}}{2} =  m \frac{v_{\rm D}^2}{2}~,
\end{equation}
where $m=\epsilon_{\rm F}/v_{\rm D}^2$ for single-layer graphene.

To conclude, we stress that the assumptions used in this parameter-matching procedure are valid thanks to the fact that all simulations 
taken into consideration in this paper work in a \textit{quasi-incompressible} regime.


%
\section{Identifying the crossover between laminar and (pre-)turbulent flow}
%



\begin{figure}[htb]
\centering
\includegraphics[width=.48\textwidth]{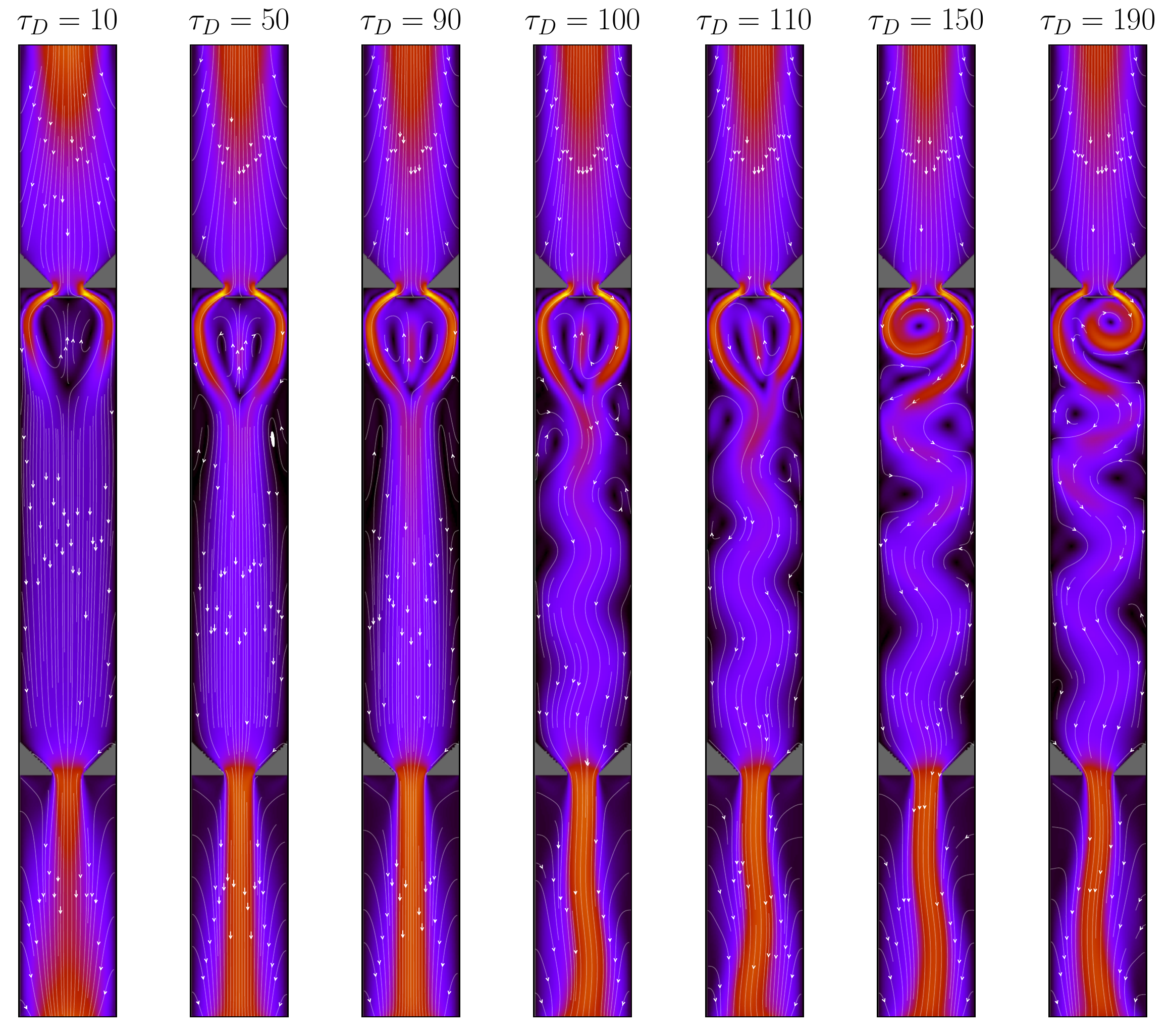} 
\includegraphics[width=.48\textwidth]{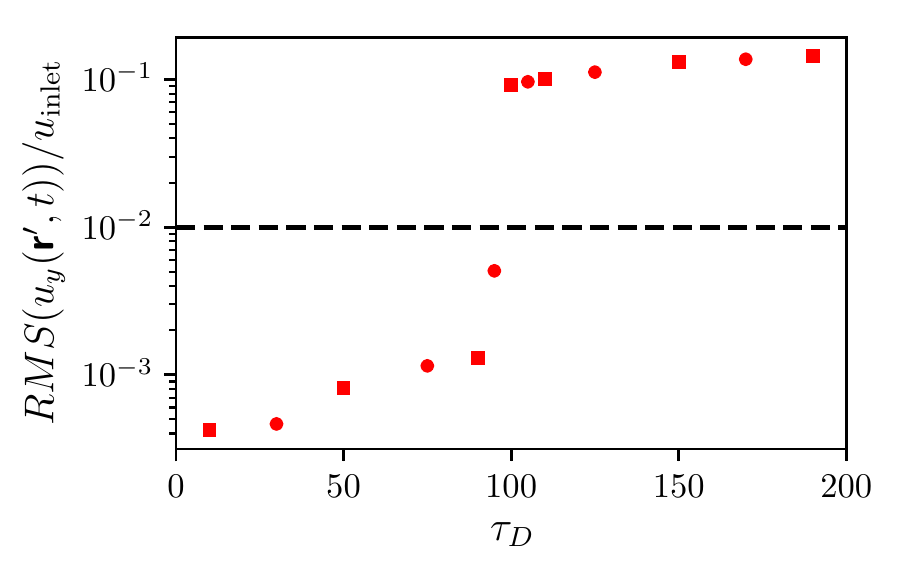} 
\caption{ The plot at right shows  the
          ratio between the root mean square transversal velocity along the middle section ($\bm{r^{\prime}}$) of the device
          and the velocity at the inlet, as a function of $\tau_D$. The black ticked line represents the (empirical) threshold used 
          to establish the crossover between a laminar and a pre-turbulent regime. In this example the crossover
          occurs in the $\tau_D^{*}$ interval $(90~\rm{ps},95~\rm{ps})$. All simulations use an inlet velocity consistent with $I = 5\cdot10^{-4}~{\rm A}$
          and $\nu = 0.45 \cdot 10^{-3} \rm{m^2/s}$. Squares refer to simulations for which a snapshot 
          of the velocity profile is shown at left; in those profiles we show the velocity streamlines with 
          colors mapping the module of the velocity.
        }\label{fig:transition_rms_velocity}
\end{figure}

In Figure 3 of the main text we show, at different values of the kinematic viscosity $\nu$,
small intervals for the  value of $\tau_D^{*}$ for which a crossover from a laminar to
a pre-turbulent flow occurs. In order to determine such intervals we have used as a discriminating factor
the onset of a transversal current ($u_y$) across the middle section of the device.
For a given simulation, we have measured at each time step the average value of $< u_y(x=L/2, y) >$. 
We consider the simulated flow  to be in a pre-turbulent regime whenever the root 
mean square of that quantity is larger than $1 \%$ of the velocity at the inlet.
In Fig.~\ref{fig:transition_rms_velocity} we show an example: the left panel  shows, in a qualitative way, the onset of pre-turbulent features in the flow
as $\tau_D$ is increased; the right panel on the other hand shows the behavior of the root mean square of $< u_y(x=L/2, y) >$ as a function of $\tau_D$.
For this particular example, we see that the crossover occurs in the $\tau_D^{*}$ interval $(90~\rm{ps},95~\rm{ps})$.

\begin{figure}[htb]
\centering
\includegraphics[width=.48\textwidth]{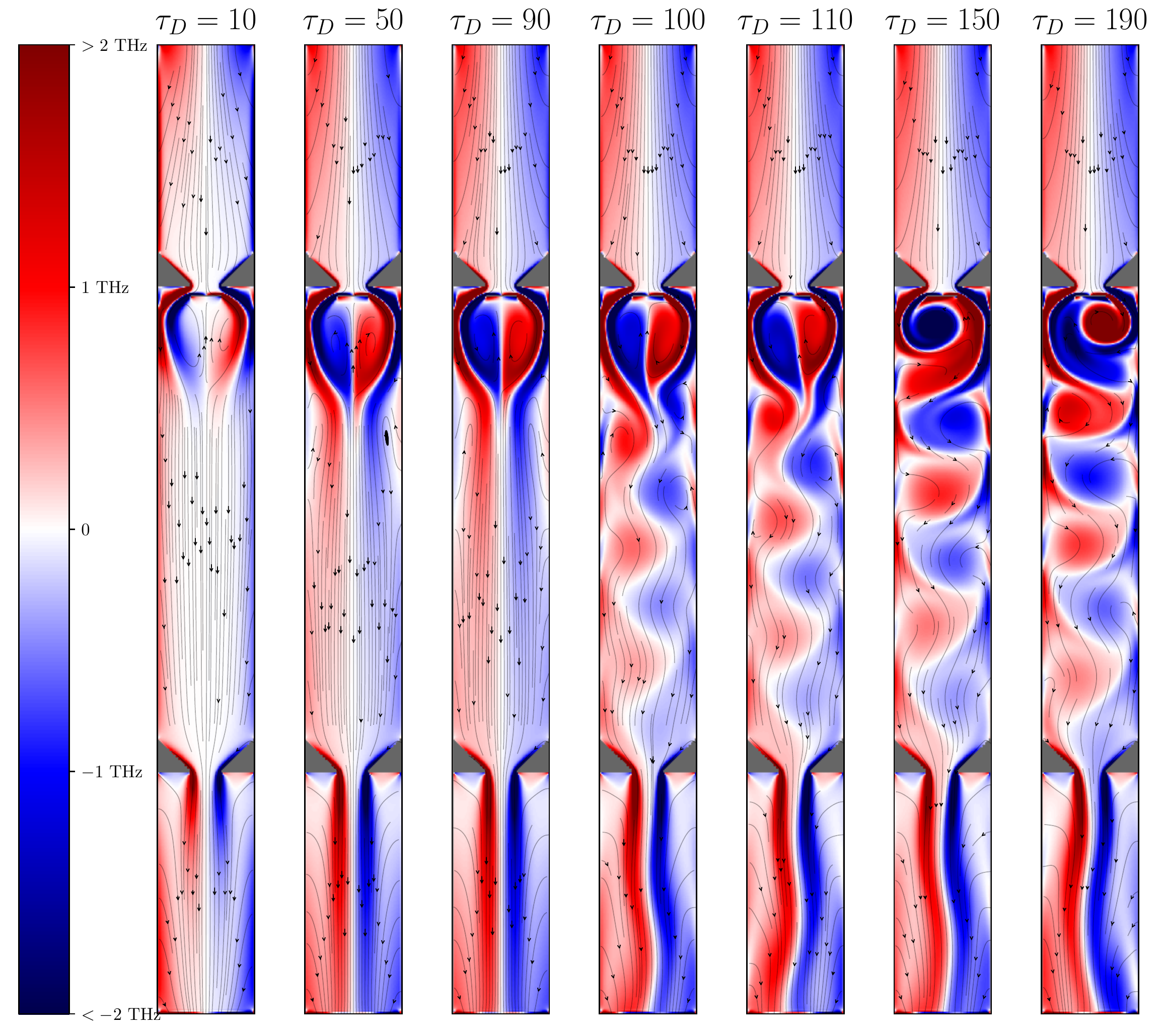} 
\includegraphics[width=.48\textwidth]{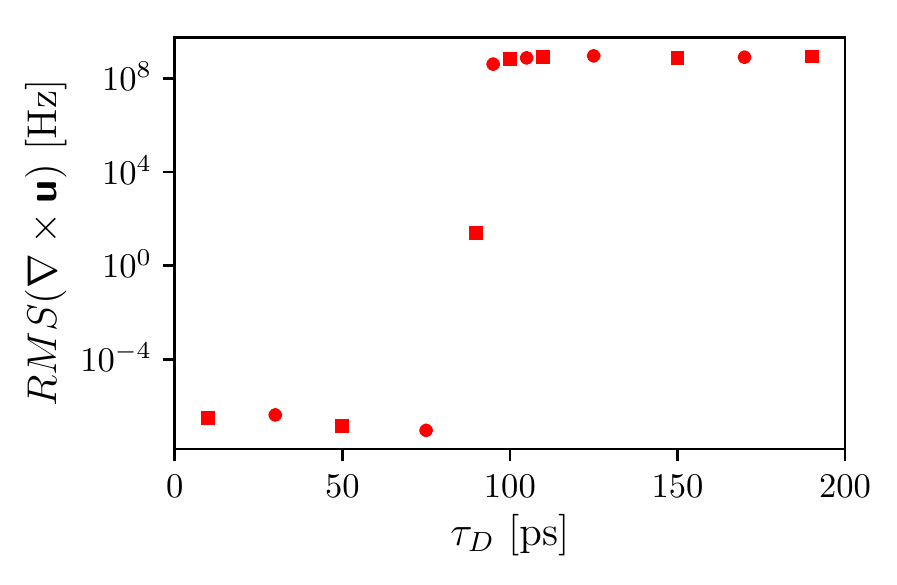}

\caption{The plot at right shows the root mean square value of the average of vorticity as a function of $\tau_D$; we once again observe an abrupt change in the $\tau_D^{*}$ interval $(90~\rm{ps},95~\rm{ps})$ similarly to Fig.~\ref{fig:transition_rms_velocity}. All simulations use the same physical parameters as in the previous figure. Squares refer 
          to simulations for which a snapshot of the vorticity profile is shown in the left panel; 
          these profiles show again the velocity streamlines but this time colors map the vorticity profile.
        }\label{fig:transition_rms_vorticity}
\end{figure}

A different criteria that could be employed to quantify the crossover 
from a laminar to a (pre)-turbulent regime consists in taking into account the vorticity,
generally defined as the curl of the velocity (a scalar in the 2D case). In particular, we 
take into consideration the root mean square (RMS) of the average value of the vorticity.
From Fig.~\ref{fig:transition_rms_vorticity} we can see that for $\tau_D^{*} < 90$ the 
average value of the vorticity is very close to zero, due to the symmetric behavior of the
laminar flow; an abrupt change occurs in the interval $\tau_D^{*} \in (90~\rm{ps},95~\rm{ps})$,
where the RMS of the average value of the vorticity grows of 6-7 orders of magnitudes. We remark that both methods yield very similar results.

\end{document}